\documentstyle[12pt,graphics]{article} 
\newcommand{\ra}{\rightarrow}
\newcommand{\bra}{\langle} 
 
\newcommand{\ket}{\rangle}

\newcommand{\be}{\begin{equation}} 
\newcommand{\ee}{\end{equation}}
\newcommand{\bea}{\begin{eqnarray}} 
\newcommand{\eea}{\end{eqnarray}}
\newcommand{\eps}{\varepsilon} 

\newcommand{\ffi}{\varphi} 
 
\newcommand{\ep}{\hfill {$\Box$}}
 
\newtheorem{lem}{Lemma}[section]
\newtheorem{prop}{Proposition}[section]
\newtheorem{thm}{Theorem}[section]

\newtheorem{asser}{Assertion}

\catcode`\@=11
\newbox\notebox%
\newdimen\leftnotesize \leftnotesize=16mm
\newskip\leftnotemargin \leftnotemargin=4mm
\newdimen\rightnotesize \rightnotesize=16mm
\newskip\rightnotemargin \rightnotemargin=4mm
\newskip\sidenotebaselineskip \sidenotebaselineskip=7\p@
\newskip\sidenoteleftskip  \sidenoteleftskip=\z@skip
\newskip\sidenoterightskip \sidenoterightskip=\z@ plus 4mm 

\def\leftnote#1{\leavevmode\vadjust{\setbox\notebox\vtop{%
   \hsize=\leftnotesize\parindent=\z@\baselineskip=\sidenotebaselineskip
   \leftskip=\sidenoteleftskip\rightskip=\sidenoterightskip\relax #1}
   \hbox{\kern-\leftnotesize\kern-\leftnotemargin\smash{\raise.5ex\box\notebox}}}}
\def\rightnote#1{\leavevmode\vadjust{\setbox\notebox\vtop{%
   \hsize=\rightnotesize\parindent=\z@\baselineskip=\sidenotebaselineskip
   \leftskip=\sidenoteleftskip\rightskip=\sidenoterightskip\relax #1}
   \line{\hfill\rlap{\smash{\kern\rightnotemargin\raise.5ex\box\notebox}}}}}

\catcode`\@=12

\let\behave\sim
\def\RL{{\cal R}_L}
\def\Dom{{\cal D}}
\def\cH{{\cal H}}
\def\cK{{\cal K}}
\def\cL{{\cal D}}
\def\gt{g_t}
\def\gs{g_s}
\def\Keps{K(\eps)}
\def\Kmeps{K_{\scriptscriptstyle-}(\eps)}
\def\Kpeps{K_{\scriptscriptstyle+}(\eps)}
\def\Kpmeps{K_{\scriptscriptstyle\pm}(\eps)}
\let\zo=o
\let\bd=u
\let\ei=r
\def\gmm{w_{*}}
\def\dist{\mathop{\rm dist}\nolimits}
\def\dom{\mathop{\rm dom}\nolimits}
\def\diag{\mathop{\rm diag}\nolimits}
\let\vx\wp
\let\vy\vartheta
\def\deriv#1{\partial#1}
\def\Mu{M}

\catcode`\@=11
\def\Bbb#1{\mbox{\bf#1}}

\def\R{{\Bbb R}} 

\def\un{\mbox{\bf I}} 
\def\N{{\Bbb N}}
\def\Z{{\Bbb Z}}
\def\Sb{{\Bbb S}}

\def\loadmsbm{%
  \font\tenmsb=msbm10 scaled\magstep1
  \font\sevenmsb=msbm7 scaled\magstep1
  \font\fivemsb=msbm5 scaled\magstep1
  \alloc@8\fam\chardef\sixt@@n\msbfam
    \textfont\msbfam=\tenmsb
    \scriptfont\msbfam=\sevenmsb
    \scriptscriptfont\msbfam=\fivemsb
  \global\let\Bbbfam\msbfam
  \gdef\Bbb##1{{\fam\msbfam\relax##1}}
  \global\def\un{{\mathchoice {\rm 1\mskip-4mu l}   {\rm 1\mskip-4mu l}
                       {\rm 1\mskip-4.5mu l} {\rm 1\mskip-5mu l}}}
  \global\let\loadmsbm\empty}

\loadmsbm
\catcode`\@=12

\begin{document}

\title{Adiabatic Evolution for Systems with Infinitely many
       Eigenvalue Crossings}
\author{A. Joye\thanks{Institut Fourier, Unit\'e Mixte de Recherche 
CNRS-UJF 5582, Universit\'e de Grenoble I, BP 74, 38402 Saint
Martin d'H\`eres Cedex, France} ,
F. Monti\thanks{Laboratoire de Physique, CNRS, Universit\'e de
Bourgogne, BP 400, 21011 Dijon, France}\ \thanks{\'Ecole Polytechnique
F\'ed\'erale de Lausanne, 1015 Lausanne, Switzerland} ,
S. Gu\'erin$\,^{\dag}$,
H.R. Jauslin$\,^{\dag}$}
\maketitle 


\begin{abstract}
We formulate an adiabatic theorem adapted to models that present an
instantaneous eigenvalue experiencing an infinite number of 
crossings with the rest of the spectrum.
We give an upper bound on the leading correction terms with respect to
the adiabatic limit. The result requires only differentiability of the
considered spectral projector, and some geometric hypothesis
on the local behaviour of the eigenvalues at the crossings.
\end{abstract}
{PACS: 03.65.Db, 03.65.Ge, 32.80.Qk, 32.80.Bx}

\vfill\eject
\section{Introduction}

The availability of intense pulsed laser sources has opened a large
field of possibilities to control atomic and molecular dynamical
processes. One of the main
theoretical tools to analyze these processes is adiabatic Floquet
theory \cite{gmdj} and references therein. The general setup can be
described as follows. One considers a  molecule described by a
Hamiltonian $H_0$ acting on a Hilbert space $\cH$, in interaction with
one radiation mode of frequency $\omega$. (The description of the
interaction with several modes of different frequencies can be
formulated along similar lines). Since the intensity of the
field is quite large, the field is treated as a classical field. The
Hamiltonian of the molecule perturbed by the electromagnetic field can
be written for example as 
\be
\label{hsc}
 H= H_0
+ E\,\Mu\,F(\omega\,t + \theta_0)
\ee
where $\Mu$ is the dipole moment of the molecule and $E\in\R$ is
the amplitude of the radiation field, $F$ is a $2\pi$-periodic function 
and $\theta_0$ the initial phase.
We assume that $H_0$ has a discrete spectrum. In order to describe a
laser pulse the amplitude is taken as a slowly varying time dependent
function $E(\varepsilon\,t)$, where one takes e.g.
$\varepsilon=1/T_p$ with $T_p$ the duration of the pulse. A new
technique that provides an efficient method for complete transfer of
population is based on frequency chirping: within the pulse duration
the frequency is also slowly modulated $\omega=\omega(\varepsilon\,t)$.

This model has  thus two kinds of time-dependencies in the  
Hamiltonian: one that is periodic and another one that is slowly
varying. The periodic part can be treated by Floquet methods, and the
slowly varying part by adiabatic theory. Adiabatic Floquet theory 
is based on the following statement: Assume that in the Hamiltonian
(\ref{hsc}) the parameter $E$ and the frequency $\omega$ are made
time dependent, $E(t)$, $\omega(t)$. Consider  the propagator
$U(t,t_0;\theta_0)$, solution of the Schr\"odinger equation 
\be\label{hu}
i\,{\partial\over \partial t} U(t,t_0;\theta_0)=
 H(\omega\,t+\theta_0)~U(t,t_0;\theta_0), \qquad U(t,t;\theta_0)=\un
\ee
acting on the Hilbert space $\cH$. We consider an enlarged Hilbert
space by tensoring $\cH$ with the space of square integrable functions
on the unit circle: $L^2(\Sb^1,\cH)$. The
operator $U(t,t_0;\theta)$ can be lifted into the enlarged space,
interpreting the $\theta$-dependence as a
multiplication operator. We can then define
$$
U_K(t,t_0)= e^{-t\omega(t)\partial}\,U(t,t_0;\theta)\,
e^{t_0\omega(t_0)\partial}
$$
where $\partial={\partial/\partial\theta}$. The statement is that
equation (\ref{hu}) is
equivalent to
\be\label{Keq}
i\,{\partial\over \partial t} U_K(t,t_0)= K(t)\,U_K(t,t_0)
\ee
with 
$$
K(t)=-i\,\varpi(t)\,{\partial \over \partial \theta} +H_0
+ E(t)\,\Mu\,F(\theta)
$$
and $\varpi(t)$ denotes an effective instantaneous frequency defined
by $\varpi(t)=\omega(t)+t\,d\omega(t)/dt$.
Assuming that the time dependence of $E(t)$, $\omega(t)$ is slow
one can develop adiabatic techniques for the evolution of (\ref{Keq}).
When $K$ has pure point spectrum, the first ingredient are the
instantaneous eigenvalues and eigenvectors. They always can be written
and labelled in the form
\be\label{struc}
\lambda_{j,k}=\lambda_{j,0}+\varpi\,k, \qquad k\in\Z
\ee
$$
  \psi_{j,k}(\theta)=\psi_{j,0}(\theta)\,e^{i\,k\,\theta}
$$
The index $j$ has the same cardinality as the dimension of the Hilbert
space $\cH$. Thus, even if we take simple models with finite dimensional
$\cH$, the Floquet spectrum has infinitely many eigenvalues. 
As functions of  $E$ and $ \varpi$, these eigenvalues may
exhibit crossings, which the adiabatic approximation can accomodate
in case there is a finite number of them, see \cite{bf}, \cite{ha0}.
The structure (\ref{struc}) of the eigenvalues is such that
if we consider
a slowly varying frequency $\varpi(t)$ that goes through $0$, one can 
encounter situations in which  a branch of instantaneous eigenvalues 
undergoes an infinite number of crossings with other branches. 
Let us stress here that this situation is not generic, as actual
crossings of eigenvalues are more the exception than the rule.
However, we give below a whole class of systems for which this
situation is true. Note also that in case $\varpi(t)$ passes through
$0$, the domain of $K(t)$
becomes time dependent, so that technical issues regarding regularity
of the evolution operator have to be addressed. This is done in the
Appendix.

The goal of the present paper is to formulate an adiabatic theorem that can 
be applied to such situations  with an estimate on the corrections to the
adiabatic limit. Adiabatic Theorems without gap conditions are known
to be true, see \cite{ae}, however, in general, no estimates on the
error terms are available.
 
While this work was motivated by the physical situation described
above and discussed below in the examples, our analysis of the
adiabatic approximation is model independent and can be applied to 
more general situations.

\section{Adiabatic Theorem}
\subsection{Context}

The adiabatic approximation in Quantum Mechanics has a long
history which we will not attempt to retrace here. We refer the reader to 
the recent surveys \cite{ae2}, \cite{jp} and references therein.
Let us simply recall here  that the works following that of Born and Fock
\cite{bf} by Kato \cite{ka1}, Nenciu \cite{ne} and Avron, Seiler, Yaffe 
\cite{asy} have lead to a formulation of the Adiabatic Theorem under 
the usual gap assuption that is general and where the error term is
well controlled and of order $\eps$. In case the gap assumption is 
modified, the situation is less explicit.
In this section, we switch back to the notation $H(\eps\,t)$ for
the slowly varying time-dependent Hamiltonian. 
Assume $H(s)$ is smooth in $s\in[0,1]$ and there exists a 
spectral projector $P(s)$ of $H(s)$ which is strongly $C^2$ on
$[0,1]$.
Avron and Elgart have shown in \cite{ae} that the adiabatic theorem holds 
under these conditions, provided $P(s)$ is of finite rank, 
independently of any spectral considerations.
The limitation of this approach is
that, in general, no estimate can be made on the rate at which the
adiabatic regime is attained. In certain specific situations, an
estimate on this rate is available. In the case where the spectral
measure $\mu_{\ffi}$ is $\alpha$-H\"older continuous, with $\ffi=
P'(s)\psi$, $\psi$ being the initial condition, the rate of
convergence was shown in \cite{ae} to be of order
$\eps^{\alpha/(2+\alpha)}$.
A case where the spectrum of $H(s)$ is assumed to be dense pure point 
is dealt with in \cite{ahs}. Another situation, considered in
\cite{j2}, where the
gap hypothesis is not necessarily fulfilled occurs when 
$H(s)=H_0(s)+\eps H_1(s)$, where the domain of $H_1(s)$ is smaller
than that of $H_0(s)$. In both cases, the error term remains of order
$\eps$. In the present article, we consider another situation in which
the usual gap assumption is modified and
the error made in the adiabatic approximation can be estimated. We
make the hypothesis that
the spectral projector $P(s)$ is associated with an eigenvalue 
$\lambda(s)$, in the sense that $H(s)P(s)=\lambda(s)P(s)$, for all
$s\in[0,1]$. We assume that  $\lambda(s)$ is isolated in the spectrum
except at a series of times $\{\zo_k\}_{k\in \N}$ accumulating at 
$a\in]0,1[$ where it experiences crossings with the rest of the
spectrum. Requiring some conditions on the local behaviour of the gap between
$\lambda(s)$ and the rest of the spectrum near the crossing points $\zo_k$, 
we estimate the error term in the theorem  without {\it a priori} 
knowledge on the nature of the rest of the  spectrum. 
Note that for $s=\zo_k$ such that $\lambda(\zo_k)$ is not isolated
in the spectrum, $P(\zo_k)$ does not represent the entire spectral projector 
associated with the eigenvalue $\lambda(\zo_k)$.

\subsection{One crossing}

Let us make more precise the regularity hypotheses under which we shall work.
In order to deal with the application described above, we will assume 
the hamiltonian is unbounded. This causes technical difficulties motivating
the part ii) of the hypothesis below which justifies our
manipulations. We show in the appendix that
this assumption is verified for our models. In case $H(s)$ is bounded, 
this part of the assumption is automatically verified.
 
\begin{itemize}
\item[$H0)$] i) We assume that for all $s\in[0,1]\backslash{\{a\}}$,
  $H(s)$ is a strongly $C^1$ self-adjoint operator defined on a 
dense domain $\Dom$ independent of $s$ 
in a separable Hilbert space ${\cal K}$, where $0<a<1$. Whereas
$H(a)$ is bounded self-adjoint on ${\cal K}$.  We also assume  
the existence of a spectral projector $P(s)$ of $H(s)$ which is  
strongly $C^2$  on $[0,1]$ and such that 
$H(s)P(s)=P(s)H(s)=\lambda(s)P(s)$, for all 
$s\in[0,1]$. \\ \\
ii)  Further assume that the unitary evolution operators 
$U(s)=U(s,0)$ and $A(s)=A(s,0)$ generated by  $H(s)$, respectively
$H(s)+\eps i[P'(s),P(s)]$ (see (\ref{evolop1}), (\ref{evolop2})) are  
well defined for all 
$s\in[0,1]$ and possess the properties i) to v) listed in Theorem 
\ref{Kato}.
\end{itemize}

Note that $P(s)$ needs not be finite dimensional and $\lambda$ is continuous.

We start by considering one crossing of $\lambda$ with
the rest of the spectrum by revisiting the strategy proposed in \cite{bf},
making use of the general analysis presented in \cite{asy}.

Let $g(s)$ be the gap between $\lambda(s)$ and the rest of the spectrum
of $H(s)$: $g(s)=\dist\big(\lambda(s),\sigma(s)
\backslash{\{ \lambda (s) \}}\big)\geq 0$, $s\in [0,1]$.
We also introduce the bounded, strongly $C^1$ operator
$L(s)=i\,[P'(s),P(s)]$.
We assume that $g^{-1}\{ 0\}=\{\zo\}$ and consider the strong differential 
equations on $\Dom$
\bea\label{evolop1}
& &i\eps\,U'(s) =H(s)\,U(s)\,,\phantom{\big(+\eps L(s)\big)}
    \qquad U(0)=\un\phantom{\bigg)} \\ \label{evolop2}
& &i\eps\,A'(s) =\big(H(s)+\eps L(s)\big)\,A(s)\,, \qquad A(0)=\un .
\eea
The unitary $A$ is the so called {\it adiabatic evolution\/} which 
possesses the well known intertwining relation $A(s)P(0)=P(s)A(s)$
\cite{ka1}, \cite{kr}.
Finally, let $W(s)$ be defined by $W(s)=A^{-1}(s)U(s)$. We have on $\Dom$
\be\label{Wdiffeq}
  iW'(s)=-A^{-1}(s)L(s)A(s)\,W(s), \qquad W(0)=\un ,
\ee
in the strong sense.
To compare the adiabatic and actual evolutions, we need to compute
the size of the difference of the unitary $W(s)$ at two times
surrounding the crossing. This is the aim of the next result.
\begin{lem}\label{TechLem1}
Under the above assumptions, we have for any
$0\leq \bd_0\leq t<\zo<s\leq \bd_1\leq 1$,
\be
 \|W(\bd_0)-W(\bd_1)\|\le C\,\left(\eps|u_0-t|/\gt^2+\eps|u_1-s|/\gs^2
  +\eps/g_t+\eps/g_s+|s-t| \right)
\ee
where $\gt=\inf_{u\in[\bd_0,t]}g(u)$, $\gs=\inf_{u\in[s,\bd_1]}g(u)$
and the constant $C$ is uniform in $\bd_0$, $\bd_1$, $s$ and $t$.
\end{lem}

\begin{figure}[h]
\hfil\scalebox{1.2}{\includegraphics{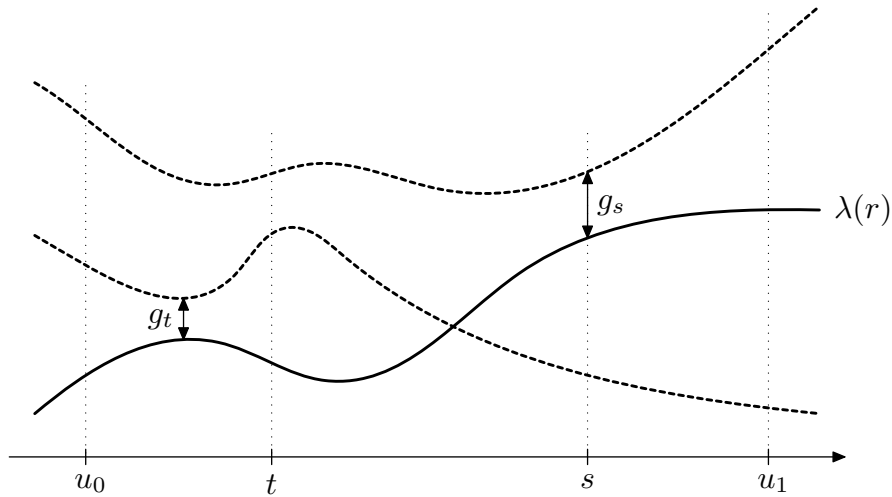}}\hfil
\caption{The various quantities defined in Lemma \ref{TechLem1}}
\end{figure}

\noindent{\bf Remark:}\\
On the basis of the classical paper by Born and Fock, \cite{bf}, and
the detailed analysis of crossings by Hagedorn \cite{ha0}, one
would expect the corresponding estimate without the first two terms.
However, such an estimate requires more detailed knowledge of the
structure of spectrum, e.g. that the gap is given by the distance 
between two eigenvalues, than what we assume in our general setting.

\medskip\noindent
Lemma \ref{TechLem1} can be used to treat two standard situations:
\begin{itemize}
\item[\it 1)] If there is a gap $G$ between $\lambda(s)$ and the rest
of the spectrum, this lemma implies that the adiabatic approximation
holds with and error term bounded by $C\,\eps/G^2$.
\item[\it 2)] If one starts the evolution on a crossing point which
splits like
$s^\alpha$ near $0$, we can use this lemma to show that the adiabatic
approximation is valid with an error bounded by
$$
\|U(1)-A(1)\|\le C\,\eps^{1/(1+2\alpha)}
$$
if $\eps$ is small enough. This is precisely the situation encountered
at the beginning of the interaction of a laser pulse with frequency
that is in resonance with the difference between two energy levels of
the molecule \cite{hol, gj}.

To get this estimate, we can consider only half of the problem by
letting aside all the terms containing a $t$ and setting $u_1=1$:
\be\label{halfcross}
 \|W(1)-W(0)\|\le C\,\left(\eps |1-s|/\gs^2+\eps/g_s+s \right)
\ee
This is indeed fully justified by the proof of the lemma (see below).
Next, we have by hypothesis that $g(s)\ge\gs=G\,s^\alpha$ if $s$ is
small. Introducing this behaviour in Equation (\ref{halfcross}), we 
obtain $\|W(1)-W(0)\|\le C\,\left(\eps/s^{2\alpha}+s\right)$. The
result follows now by balancing the two contributions by choosing
$s=s(\eps)=\eps^{1/(1+2\alpha)}$. Again, with more information
on the spectrum, as in \cite{bf, ha0}, one should be able to improve 
the above estimate to order $ \eps^{1/(1+\alpha)}$.
\end{itemize}
{\bf Proof of Lemma \ref{TechLem1}:}\\
The idea of the proof is to integrate Equation (\ref{Wdiffeq}) over 
the interval $[\bd_0,\bd_1]$ and then to get ``nice'' estimates of 
the sizes on each subintervals $[\bd_0,t]$, $[t,s]$ and $[s,\bd_1]$.
By integrating Equation (\ref{Wdiffeq}), we get
\bea
& &\displaystyle i\,\big(W(\bd_1)-W(\bd_0)\big)=
-\int_{\bd_0}^{t}A^{-1}(u)L(u)A(u)\,W(u)\,du\\
 & &\qquad\qquad\displaystyle-\int_{t}^{s}A^{-1}(u)L(u)A(u)\,W(u)\,du
-\int_{s}^{\bd_1}A^{-1}(u)L(u)A(u)\,W(u)\,du\nonumber
\eea
For the middle term, we simply use the properties of the operator
norm and the fact that $A(u)$ and $W(u)$ are unitary to obtain:
\be\label{est1}
\big\|W(s)-W(t)\big\|\le\int_t^s\big\|L(u)\big\|\,du\le
\sup_{u\in[0,1]}\big\|L(u)\big\|\,|s-t|\,,
\ee
i.e.\ we do not care about the behaviour of $g(u)$ inside the
subinterval $[t,s]$.\\
To estimate the first integral, let $Q(u)=\un-P(u)$. A simple
computation, using $P(s)\,P'(s)\,P(s)\equiv0$, shows that
\be\label{diagterms}
P(u)\,L(u)\,P(u)=Q(u)\,L(u)\,Q(u)=0\,,
\ee
and due to the intertwining property of $A(u)$, we can write
\bea\label{firstint}
& &W(t)-W(\bd_0)
  =i\int_{\bd_0}^{t}\Big(P(0)A^{-1}(u)L(u)A(u)Q(0)\nonumber\\
& &\hphantom{W(t)-W(\bd_0)=\int_{\bd_0}^{t}\Big(P(0)A}
+Q(0)A^{-1}(u)L(u)A(u)P(0)\Big)W(u)\,du
\eea
Now, we need to extract an explicit $\eps$ dependence from this
equality in order to obtain the estimates stated in the lemma. To do 
this, we follow \cite{asy} and introduce the bounded operator 
$\RL(u)$ defined by
$$
\RL(u)=
\frac{1}{2i\pi}\oint_{\Gamma(u)}R(u,\lambda)L(u)R(u,\lambda)\,d\lambda
$$
where $R(u,\lambda)=\Big(H(u)-\lambda\Big)^{-1}$ is the resolvent of
$H(u)$ at $\lambda$ and where the loop $\Gamma(u)$ is a circle centered
at $\lambda(u)$ of radius $g(u)/2$.
It has the properties (see \cite{asy}, \cite{j2})
\bea\label{proprl}
  & &[\RL(u),H(u)]=[L(u),P(u)]\\
  & &P(u)\RL(u) P(u)=Q(u)\RL(u) Q(u)=0.
\eea\\
Standard arguments show that $\RL(u)$ is strongly
$C^1$ and that
\bea
& &\RL'(u)=
   \frac{1}{2i\pi}\oint_{\Gamma(u)}\Big(R(u,\lambda)L'(u)R(u,\lambda)
   \nonumber\\
  & &\hphantom{\RL'(u)=\frac{1}{2i\pi}\oint_{\Gamma(u)}\Big(R(u,}
    -R(u,\lambda)H'(u)R(u,\lambda)L(u)R(u,\lambda) \\
    & &\hphantom{\RL'(u)=\frac{1}{2i\pi}\oint_{\Gamma(u)}\Big(R(u,}
     -R(u,\lambda)L(u)R(u,\lambda)H'(u)R(u,\lambda)\Big)
   d\lambda\,,\nonumber
\eea
where $H'(u)R(u,\lambda)$ is to be understood as the bounded operator
\be\label{hprim}
 H'(u)R(u,\lambda)=H'(u)R(u,i)\,\big(\un+(\lambda-i)R(u,\lambda)\big).
\ee
Hence, we get the following estimates:
\bea\label{esr}
  \big\|\RL(u)\big\|&\leq&{|\Gamma(u)|\over2\pi}\,\big\|L(u)\big\|
  \,\big(g(u)/2\big)^{-2}=2\big\|L(u)\big\|/g(u).\phantom{\bigg)}\\
\label{esrprim}
  \big\|\RL'(u)\big\|&\leq&
  c\max\Big\{\big\|H'(u)R(u,i)\big\|\,\big\|L(u)\big\|,
             \big\|L'(u)\big\|\Big\}/g^2(u).
\eea
The main property of $\RL(u)$ (see \cite{asy}) is that it satisfies
for any $\psi\in\Dom$ the following equalities, as verified by means
of (\ref{proprl}):
\bea\label{eqLP}
& &P(0)A^{-1}(u)L(u)A(u)Q(0)\,\psi=
  -i\eps\,\frac{d}{du}\left(P(0)A^{-1}(u)\RL(u)A(u)Q(0)\,
  \psi\right)\nonumber\\
& &\phantom{Q(0)A^{-1}(u)L(u)A(}+i\eps\,P(0)A^{-1}(u)\RL'(u)A(u)Q(0)\,
\psi\,.
\eea
and
\bea\label{eqLQ}
& &Q(0)A^{-1}(u)L(u)A(u)P(0)\,\psi=
  i\eps\,\frac{d}{du}\left(Q(0)A^{-1}(u)\RL(u)A(u)P(0)\,
  \psi\right)\nonumber\\
& &\phantom{Q(0)A^{-1}(u)L(u)A(}-i\eps\,Q(0)A^{-1}(u)\RL'(u)A(u)P(0)\,
\psi\,.
\eea
These equations imply that $\int_{u_0}^tA^{-1}(u)L(u)A(u)W(u)du$ is 
proportional
to $\eps$. Indeed, Equalities (\ref{diagterms}) and the
intertwining property of $A(u)$ show that the diagonal blocks are
$0$.\\
Introducing Equalities (\ref{eqLP}) and (\ref{eqLQ}) in Equation 
(\ref{firstint}), we get
\bea
\!\!\!\!& &W(t)-W(\bd_0)=\\
& &\quad 
  -\eps\int_{\bd_0}^t\!\frac{d}{du}\!\Big(Q(0)A^{-1}(u)\RL(u)A(u)P(0)
     \!-\!P(0)A^{-1}(u)\RL(u)A(u)Q(0)\Big)W(u)du\nonumber\\
& &\quad
  -\eps\int_{\bd_0}^t\!\Big(P(0)A^{-1}(u)\RL'(u)A(u)Q(0)
     \!-\!Q(0)A^{-1}(u)\RL'(u)A(u)P(0)\Big)W(u)du\,.\nonumber
\eea
Performing an integration by part in the first integral, using the
differential equation (\ref{Wdiffeq}) for $W(u)$ and taking into
account that $A(u)$, $W(u)$ are unitary and $P(0)$, $Q(0)$ are
projectors, gives us the following bound for the norm of the 
difference $W(t)-W(\bd_0)$:
\bea
& &\big\|W(t)-W(\bd_0)\big\|\le
2\eps\,\Big(\big\|\RL(t)\big\|+\big\|\RL(\bd_0)\big\|\nonumber\\
& &\qquad\quad+\sup_{u\in[\bd_0,t]}\big\|\RL(u)\big\|\,\big\|L(u)\big\|
\,(t-\bd_0)+\sup_{u\in[\bd_0,t]}\big\|\RL'(u)\big\|\,(t-\bd_0)\Big)\,.
\eea
Next, we use first Estimates (\ref{esr}) and (\ref{esrprim}) and 
then the fact that $0\le \bd_0<t\le1$ to obtain the desired bound:
\bea\label{est2}
\big\|W(t)-W(\bd_0)\big\|&\!\le\!&
{8\eps\over\gt}\,\sup_{u\in[\bd_0,t]}\big\|L(u)\big\|
+{4\eps\over\gt}\,\sup_{u\in[\bd_0,t]}\big\|L(u)\big\|^2\,(t-\bd_0)
\nonumber\\
& &\qquad+c\,{2\eps\over\gt^2}\,\sup_{u\in[\bd_0,t]}
\Big\{\big\|H'(u)R(u,i)\big\|\,\big\|L(u)\big\|,
\big\|L'(u)\big\|\Big\}\,(t-\bd_0)\nonumber\\
&\!\le\!&
12\,{\eps\over\gt}\,\sup_{u\in[0,1]}\Big\{\big\|L(u)\big\|,
\big\|L(u)\big\|^2\Big\}\\
& &\qquad+2\,{\eps|t-u_0|\over\gt^2}\,\sup_{u\in[0,1]}
\Big\{\big\|H'(u)R(u,i)\big\|\,\big\|L(u)\big\|,
\big\|L'(u)\big\|\Big\}\nonumber\\
&\!\le\!&c_2\,\Big({\eps\over\gt}+{\eps|t-u_0|\over\gt^2}\Big)
\,.\nonumber
\eea
Using the same kind of arguments, shows that on the
subinterval $[s,\bd_1]$, we have
\be\label{est3}
\big\|W(\bd_1)-W(s)\big\|\le c_2\,\Big({\eps |s-u_1| \over\gs^2}+
{\eps \over\gs}\Big)\,.
\ee
Combining Estimates (\ref{est1}), (\ref{est2}) and (\ref{est3}) gives
the announced bound for $\big\|W(\bd_1)-W(\bd_0)\big\|$.
\ep

\subsection{Infinite number of crossings}

We now have all the information required to proceed to the case 
of an infinite number of crossings. We make the following hypotheses
describing what happens in the neighbourhood of each crossing.

\smallskip\noindent{\bf Spectral Hypotheses:}
There exist two partitions $\{\bd^{\pm}_k\}_{k\in\N}$ of $[0,a)$
and $(a,1]$ respectively:  
$$
0=\bd^-_0<\ldots<\bd^-_{k-1}<\bd^-_{k}\ldots
\longrightarrow\bd^-_\infty=a=\bd^+_\infty\longleftarrow\ldots
\bd^+_{k}<\bd^+_{k-1}<\ldots<\bd^+_0=1
$$ 
such that for each $k\in\N^*$, 
\begin{itemize}
\item[\it H1)]  one can find non empty open
  intervals $V_k^{\pm}$, which satisfy
  $V^-_k\subset[u_{k-1}^-,u_{k}^-]$,
  $V^+_k\subset[u_{k}^+,u_{k-1}^+]$ and 
  \be 
    \sup_{s\in V_k^{\pm}}g(s)\leq \inf_{t\in I_{k}^{\pm}} g(t)
  \ee 
  where $I_k^-=[u_{k-1}^-,u_{k}^-]\backslash{V^-_k}$ and 
  $I_k^+=[u_{k}^+,u_{k-1}^+]\backslash{V^+_k}$.
\item[\it H2)]  there are constants
  $G_{\pm}(k)>0$ and a $k$-independent positive constant $\alpha$
  such that for all $s\in V_k^{\pm}$:
  \be
    G_{\pm}(k)\,|s-\zo^{\pm}_k|^{\alpha}\leq g(s)\,,
  \ee
  for some points $\zo^{\pm}_k\in V_k^{\pm}$.
\end{itemize}
\begin{figure}[h]
\hfil\scalebox{1.2}{\includegraphics{figsxx.2}}\hfil
\caption{Illustration of the spectral hypotheses {\it H1--H2\/} on the
         interval $(0,a)$. The intervals $V^-_l$ are represented by 
         $\lgroup\!\!\!\lgroup\!\!\raise2.333pt
         \hbox to3em{\vrule height 1.5pt width 2.9em\hfil}
         \!\!\rgroup\!\!\!\rgroup$.}
\end{figure}

\noindent{\bf Comments:}\\
{\it 1)} These Spectral Hypotheses mean that the
crossings are well separated and that they behave as power of
order at most $\alpha$. Hypothesis {\it H1\/} tells us that
outside the crossing regions $(V_k^{\pm})$ the gaps are relatively
``large''. This means that the only accumulation point of
small gaps is $a$.\\
{\it 2)} The choice of a constant exponent $\alpha$ is not as
restrictive as it might look at first. Indeed, we are interested in
an upper bound, so it is the greatest $\alpha$ that will determine
the global behaviour.\\
{\it 3)} 
In the applications, we will consider examples where
$g^{-1}\{0\}=\{\zo^{\pm}_k\}$: the set of
crossing points of $\lambda(s)$ with the rest of the spectrum.
This implies $\alpha>0$. But, the
case of an infinite number of avoided crossings can be treated by
taking $\alpha=0$ in Hypothesis {\it H2\/}.

To obtain an estimate for the difference between the real evolution
$U(1)$ and the adiabatic one $A(1)$, the idea is to apply Lemma 
\ref{TechLem1} on a finite number of crossings and to take a simple
integral bound (as in (\ref{est1})) over the rest of the interval
surrounding $a$. The choice of the number of crossings will be 
optimized with respect to $\eps$ in order to get a simple form for
the bound of the remainder term.
To state the corresponding result, we need to introduce some
notations.
Let $\Delta_{\pm}(k)=\max\Big\{|\bd^{\pm}_k-\zo^{\pm}_k|,
|\bd^{\pm}_{k-1}-\zo^{\pm}_{k}|\Big\}$ and
$\tau_{\pm}(k)=\max\Big\{{\Delta_{\pm}(k)/G_{\pm}^2(k)},
  {\Delta^\alpha_{\pm}(k)/G_{\pm}(k)}\Big\}$.
The functions $K\mapsto
|\bd_K^\pm-a|/\sum_{k=1}^K\tau_\pm(k)^{1/(1+2\alpha)}$ are
monotonically decreasing to zero, so, if $\eps$ is small enough,
we define $\Kpmeps\in\N^*$ as the greatest integer satisfying
\be\label{defk}
  \frac{|\bd_{K}^\pm-a|}{\displaystyle\sum\nolimits_{k=1}^{K}
   {\tau_\pm(k)}^{1/(1+2\alpha)}}\ge\eps^{1/(1+2\alpha)}\,.
\ee
This integer always exists if $\eps$ is sufficiently small and, by
construction, $\Kpmeps\ra\infty$ as $\eps\ra0$.
\begin{thm}\label{adiab}
For $\eps$ small enough, under H0 and the Spectral Hypotheses H1, H2 and 
provided that
\be\label{epsGcond}
\varsigma\,\big(\eps\,\tau_{\pm}(k)\big)^{1/(1+2\alpha)}
\leq |V_k^{\pm}|/2
\qquad\mbox{for all }1\leq k\leq \Kpmeps\,,
\ee
for some constant $\varsigma>0$, we have that 
$$
U(1)=A(1)+
     O\left(\max\{|\bd^-_{\Kmeps}-a|,|\bd^+_{\Kpeps}-a|\}\right)\,.
$$
Hence, as $\lim_{\eps\ra0}\Kpmeps=\infty$, $\big\|U(1)-A(1)\big\|$
goes to zero for $\eps\ra 0$ as fast as 
$\max\{|\bd^-_{\Kmeps}-a|,|\bd^+_{\Kpeps}-a|\}$.
\end{thm}
{\bf Remarks:}\\
{\it 1)\/} The theorem states that the error can be estimated provided
we can compute the critical value $\Kpmeps$. Further considerations on
the practical aspects of this computation are given in the next
section.\\
{\it 2)\/} Condition (\ref{epsGcond}) implies that the size of the
intervals $V_k^{\pm}$ cannot be too small with respect to
$\eps\,\tau_{\pm}(k)$.\\
{\it 3)\/} While we shall apply the theorem in a situation where the
spectrum is simple and pure point, the theorem remains valid under the sole
existence of an eigenvalue separated from the rest of the spectrum by
gaps with the properties stated in H1-H2, without any knowledge on the
rest of the spectrum or restriction on the dimension of $P(s)$.

\noindent
{\bf Proof:}\\ 
In the sequel, we will denote by the same symbol $c$ all inessential
constants. Let us consider the interval $[0;a)$. In order to simplify
the notations, we will not write the sub/super-scripts $-$. Picking
some $t,s\in V_k$ such that $t<\zo_k<s$ and
$|t-\zo_k|=|s-\zo_k|$, we get
\bea\label{Wkdiff}
& &\hspace*{-2em}\|W(\bd_k)-W(\bd_{k-1})\|
\leq c\,\left(\eps |t-\bd_{k-1}|/\gt^2+\eps
   |s-\bd_k|/\gs^2+\eps/g_t +\eps/g_s+ |t-s|\right)
\nonumber\\
 & &\leq c\,
   \left(\eps\,{\Delta(k)\over G(k)^2}\,|t-\zo_k|^{-2\alpha}+
\eps\,{1\over G(k)}\,|t-\zo_k|^{-\alpha}+|t-\zo_k|\right)
\nonumber\\ \label{loose}
 & &\leq c\,
   \left(\eps\,{\Delta(k)\over G(k)^2}\,|t-\zo_k|^{-2\alpha}+
\eps\,{\Delta^\alpha(k)\over G(k)}\,
   |t-\zo_k|^{-2\alpha}+|t-\zo_k|\right)
  \\
 & &\leq c\,
   \left(\eps\,\tau(k)\>|t-\zo_k|^{-2\alpha}+|t-\zo_k|\right)
\eea
by the preceding section. 
Indeed, we have that $\gt=\inf_{u\in[\bd_{k-1},t]}g(u)=g(r_t)$ for
some $r_t\in[\bd_{k-1},t]$. Now, by Hypothesis H1,
$r_t\in V_k$. Whence, we have that
$$
\gt=g(r_t)\ge G(k)\,|r_t-\zo_k|^\alpha
\ge G(k)\,|t-\zo_k|^\alpha
$$
as $r_t\le t\le \zo_k$. Using the same kind of arguments, we can show
that $\gs=\inf_{u\in[s,\bd_{k}]}g(u)\ge G(k)\,
|s-\zo_k|^\alpha$. Finally to obtain the bound (\ref{Wkdiff}), it
remains to notice that $|s-t|=|t-\zo_k|+|s-\zo_k|=2\,|t-\zo_k|$
together with $|t-\zo_k|$, $|t-\bd_{k -1}|\leq \Delta(k)$
and  $|s-\zo_k|$, $|s-\bd_k|\leq \Delta(k)$.

We now get an estimate by choosing $t=t(\eps,k)$ in order to 
balance the two contributions appearing in the last term of 
Equation (\ref{Wkdiff}) above: for some constant $\varsigma>0$, 
we set 
\be
  \frac{\varsigma^{1+2\alpha}\,\eps\,\tau(k)}
   {|t(\eps,k)-\zo_k|^{2\alpha} }=|t(\eps,k)-\zo_k|,
\ee 
i.e.
\be\label{tek}
|t(\eps,k)-\zo_k|=\varsigma\,
\Big(\eps\,{\tau(k)}\Big)^{1/(1+2\alpha)}
\,.
\ee
By definition, $t(\eps,k)\in V_k$, hence, as $k$ will eventually be
bounded from above by $K(\eps)$, this imposes Condition 
(\ref{epsGcond}) in the statement of the theorem. Replacing $t$ by
$t(\eps,k)$ in (\ref{Wkdiff}) and summing over $k$, we get for any
$K\leq K(\eps)$, 
\be\label{WKest1}
\|W(0)-W(\bd_K)\|\leq c\,(\varsigma+\varsigma^{-2\alpha})
\sum_{k=1}^K\Big(\eps\,{\tau(k)}\Big)^{1/(1+2\alpha)}\,.
\ee
On the other hand, using the differential Equation
(\ref{Wdiffeq}), we obtain,
\be\label{WKest2}
  \|W(\bd_K)-W(a)\|\le\int_{\bd_K}^{a}\|L(u)\|\,du
\leq c\,|\bd_K-a|\,.
\ee
Again, we balance the two right hand sides in (\ref{WKest1}) and
(\ref{WKest2}) by setting the integer $K=\Keps$, which has been
defined in Equation (\ref{defk}). Consequently,
\bea\label{error}
  \|W(0)-W(a)\|&\leq&c\,\bigg((\varsigma+\varsigma^{-2\alpha})\,
  \eps^{1/(1+2\alpha)}
  \sum_{k=1}^{\Keps}{\tau(k)}^{1/(1+2\alpha)}+
  |\bd_{\Keps} -a|\bigg)\nonumber\\
  &\leq&c\,(\varsigma+\varsigma^{-2\alpha}+1)\,
 |\bd_{\Keps}-a|\equiv C(\varsigma)\,|\bd_{\Keps}-a|
\eea
where $C(\varsigma)$ is independent of $\eps$. Proceeding similarly
on $(a,1]$ completes the proof. 
\ep \\
{\bf Remarks:} \\
{\it 1)\/} The introduction of an adjustable constant $\varsigma$ 
is necessary in the following application to satisfy the hypothesis of
the theorem.\\
{\it 2)\/} In the step (\ref{loose}) we deliberately lost a little in the
estimate by using $ |t-\zo_k|^{-\alpha}\leq \Delta_k^{\alpha} 
|t-\zo_k|^{-2\alpha}$ in order to simplify the subsequent arguments. It
is nevertheless possible to get slightly sharper results by not adopting this 
simplification, however the analysis gets more involved and less
transparent. 
We simply note here that in the examples discussed below, this
more careful analysis yields, for the generic situation, an error 
term of order $\epsilon^{p}$ with an exponent $p=1/3$, instead of
the value $p=1/3-\nu$, for any $\nu >0$ obtained there.

\section{Application}

We can obtain more explicit estimates on the rest by considering some
specific behaviour at the crossings.

Let us introduce the following notation: $F_k\behave f(k)$ means that
there exist two constants $0<c_1<c_2<\infty$ such that
$c_1\,f(k)\le F_k\le c_2\,f(k)$ for $k\in\N^*$ large enough. We have
the
\begin{prop}\label{behaveProp}
Assume the hypothesis of Theorem \ref{adiab} and the following 
behaviour for the relevant quantities:
$$
\begin{array}{ll}
  |\bd_{k}^\pm-a|=C_1/k^{\beta}+C_2/k^{\beta+1}+o(1/k^{\beta+1}))\,,& 
  \qquad \beta >0, C_1 \neq 0\\
  G_\pm(k)\behave k^{\gamma}\,, \vphantom{\bigg)}\\
  |V_k^\pm|\behave1/k^{\delta}\,,& \qquad \delta >0\,.
\end{array}
$$
We set $\mu=\min\{\beta+1+2\gamma,\alpha(\beta+1)+\gamma\}$.
Then $\big\|U(1)-A(1)\big\|=O(\eps^p)$ where the exponent $p$ is
given by
$$
p=\left\{\begin{array}{ll}
    \displaystyle{1\over1+2\,\alpha}&\quad\vphantom{\Bigg)}
    \mbox{if }\mu>(1+2\alpha)\cr
    \displaystyle{1\over1+2\,\alpha}-\nu\quad\forall\nu>0&\quad
    \mbox{if }\mu=(1+2\alpha)\vphantom{\Bigg)}\cr
    \displaystyle{\beta\over(\beta+1)\,(1+2\alpha)-\mu}&\quad
    \mbox{if }\mu<(1+2\alpha)\vphantom{\Bigg)}
  \end{array}\right.
$$
provided that $\delta$ satisfy the following constraints:
$\beta+1\le\delta\le\beta+\max\{1,\mu/(1+2\alpha)\}$.
\end{prop}

{\bf Proof:}\\
The idea of the proof is to explicit conditions on the different 
exponents ensuring the validity of Theorem \ref{adiab}. We will only 
consider the interval $[0,a)$, the same kind of arguments will apply 
on $(a,1]$. Again, in order to simplify the notations we will let
aside the sub/super-scripts $-$.

First, we have that 
$2\,\Delta(k)=\bd_k-\bd_{k-1}=C_1\beta/k^{\beta+1}+o(1/k^{\beta+1})
\behave1/k^{\beta+1}$, which implies that
\be\label{del}
 \delta\geq\beta+1>0,
\ee
since $2\,\Delta(k)\ge|V_k|\behave 1/k^{\delta}$. Notice that the
length of the $V_k$ can be rescaled by a uniform constant if
$\delta=\beta+1$.\\
Next, $\Delta(k)/G^2(k)\behave 1/k^{\beta+1+2\gamma}$ and
$\Delta^\alpha(k)/G(k)\behave 1/k^{\alpha(\beta+1)+\gamma}$. So, if we
denote by $\mu=\min\{\beta+1+2\gamma,\alpha(\beta+1)+\gamma\}$ then
$\tau(k)=\max\{\Delta(k)/G^2(k),\Delta^\alpha(k)/G(k)\}\behave 
1/k^{\mu}$ by increasing the overall constant in Theorem \ref{adiab}
if necessary. Whence,
\be
  \sum_{k=1}^K{\tau(k)}^{1/(1+2\alpha)}\behave
  \sum_{k=1}^K k^{-\mu/(1+2\alpha)}\behave
  \left\{\begin{array}{ll}
     K^{0}&\mbox{if }\mu>1+2\alpha\cr
     \log K&\mbox{if }\mu=1+2\alpha\phantom{\bigg)}\cr
     K^{1-\mu /(1+2\alpha)}&\mbox{if }\mu<1+2\alpha
  \end{array}\right.
\ee
and considering the definition of $\Keps$ (see Eq.~\ref{defk}), we
obtain
\be\label{epsbehave}
\eps^{1/(1+2\alpha)}\behave\frac{|\bd_{\Keps}-a|}
 {\displaystyle\sum_{k=1}^{\Keps}{\tau(k)}^{1/(1+2\alpha)}}\behave
 \left\{\begin{array}{ll}
   \Keps^{-\beta}&\mbox{if }\mu>1+2\alpha\cr
   \Keps^{-\beta}/\log\Keps&\mbox{if }\mu=1+2\alpha
   \phantom{\bigg)}\cr
   \Keps^{-\beta-1+\mu/(1+2\alpha)}&\mbox{if }\mu<1+2\alpha
 \end{array}\right.
\ee
Condition (\ref{epsGcond}) stated in Theorem \ref{adiab} reads
\be
  \varsigma\,\Big(\eps\,\tau(k)\Big)^{1/(1+2\alpha)}\leq
  |V_k|/2
\ee
for all $1\leq k\leq\Keps$. 
Notice that this condition is automatically satisfied if 
$\delta<\mu/(1+2\alpha)$. In general, it will be satisfied for
a sufficiently small $\varsigma$, if 
\be
 F(\eps)\equiv\eps^{1/(1+2\alpha)}\Keps^{\delta-\mu/(1+2\alpha)}
\ee
remains bounded as $\eps\ra 0$. Using (\ref{epsbehave}), we have
\be\label{Fbehave}
 F(\eps)\behave\left\{\begin{array}{ll}
   \Keps^{\delta-\beta-\mu/(1+2\alpha)} 
         &\mbox{if }\mu>1+2\alpha\,,\cr
   \Keps^{\delta-1-\beta}/\log\Keps&\mbox{if }\mu=1+2\alpha\,,
   \phantom{\bigg)}\cr
   \Keps^{\delta-\beta-1}&\mbox{if }\mu<1+2\alpha\,.
  \end{array}\right. 
\ee
As $\Keps\ra\infty$ for $\eps\ra0$, Equation (\ref{Fbehave}) implies
that $F(\eps)$ will remain bounded if $\delta\le\beta+
\max\{1;\mu/(1+2\alpha)\}$. 

Hence, using (\ref{error}) and (\ref{epsbehave}), we get that the 
Adiabatic Theorem \ref{adiab} holds with a remainder term on $[0,a)$,
$$
O\left(|\bd_{\Kmeps}-a|\right)=O\left(\Keps^{-\beta}\right)
 =O\left(\eps^{p}\right)
$$
where the exponent $p$ is given by
$$
p=\left\{\begin{array}{ll}
    \displaystyle{1\over1+2\,\alpha}&\quad\vphantom{\Bigg)}
    \mbox{if }\mu>(1+2\alpha)\cr
    \displaystyle{1\over1+2\,\alpha}-\nu\quad\forall\nu>0&\quad
    \mbox{if }\mu=(1+2\alpha)\vphantom{\Bigg)}\cr
    \displaystyle{\beta\over(\beta+1)\,(1+2\alpha)-\mu}&\quad
    \mbox{if }\mu<(1+2\alpha)\vphantom{\Bigg)}
  \end{array}\right.
$$
provided that $\beta+1\le\delta \le\beta+\max\{1;\mu/(1+2\alpha)\}$.
To determine $p$ in case $\mu=1+2\alpha$ and $\delta=\beta+1$,
we have used the estimate $\eps^{-1/(1+2\alpha)}\behave
\Keps^{\beta}\log\Keps<\Keps^{\beta+\nu\,'}$ for all $\nu\,'>0$.
This ends the proof of the proposition.
\ep 

\vspace{.5cm}

\noindent{\bf Remark:}\\
We can strengthen the last comment of the preceding section with
the following observation.
In case $\alpha=\beta=\gamma=1$ and $\delta=2$, this last comment 
asserts that we have $p=1/3$, instead of $p=1/3-\nu$, for all 
$\nu>0.$ Now,
if in Lemma 2.1, the right member were missing the terms 
$\eps|u_0-t|/g_t^2 +\eps|u_1-s|/g_s^2 $, as one would expect 
with a little more information on the spectrum, an analysis 
similar to the one provided above leads to an error term of 
order $\eps^{1/3}$. This makes it reasonable 
to expect that in such a situation the error actually is of that order,
as it was the case in the corresponding analysis of one crossing 
performed in \cite{bf}, see \cite{ha0}. Finally, it is shown in the
examples below that the values  $\alpha=\beta=\gamma=1$ and $\delta=2$
are generic in some sense.

\section{Examples}

We now consider a family of models for which the situation
just described takes place as the effective frequency $\varpi$ takes
the value zero. 
We start by considering the most general model for a two level system
driven by a periodic field. The model can be characterized by choosing
freely the eigenvalues $\lambda_{+,m}=\lambda_++m\,\omega$ 
and $\lambda_{-,k}=\lambda_-+k\,\omega$ and the corresponding
eigenfunctions of the form: 
\be\label{defpsi}
\psi_{+,m}(\theta)=
\pmatrix{e^{i\,x(\theta)}\,\cos z(\theta)\cr
         e^{i\,y(\theta)}\,\sin z(\theta)\cr}\,e^{i\,m\,\theta}
\qquad\mbox{and}\qquad
\psi_{-,k}(\theta)=
\pmatrix{-e^{-i\,y(\theta)}\,\sin z(\theta)\cr
         e^{-i\,x(\theta)}\,\cos z(\theta)\cr}\,e^{i\,k\,\theta}\,,
\ee
in which the functions $x$, $y$ and $z$ are periodic modulo an integer
multiple of $\theta$.

Defining the unitary matrix
$$
\Upsilon(\theta)=\pmatrix{
  e^{i\,x(\theta)}\,\cos z(\theta)&-e^{-i\,y(\theta)}\,\sin z(\theta)\cr
  e^{i\,y(\theta)}\,\sin z(\theta)
  &\phantom{-}e^{-i\,x(\theta)}\,\cos z(\theta)\cr
}
$$
the corresponding Floquet Hamiltonian can be written as (dropping the
$\theta$ dependence in the notation)
$$
K=-i\,\varpi\,\partial-i\,\varpi\,\Upsilon\,(\partial\Upsilon^{-1})+
\Upsilon\,D\,\Upsilon^{-1}
$$
where $D=\diag(\lambda_+,\lambda_-)$. Using the notation $2\vx=x+y$,
$2\vy=y-x$ and choosing, without loss of generality,
$\lambda_+=-\lambda_-=\lambda$, the Floquet Hamiltonian can be
expressed as
\bea\label{Kdef}
& &\!\!\!\!\!\!\!\!K=-i\,\varpi\partial\phantom{\Bigg)}\\
& &\!\!\!\!\!\!\!\!\phantom{K=}+
\pmatrix{\varpi\deriv\vy+(\lambda-\varpi\deriv\vx)\cos(2z)&
\Big(-i\varpi\deriv{z}+(\lambda-\varpi\deriv\vx)\sin(2z)\Big)
e^{-2i\vy}\cr
\Big(i\varpi\deriv{z}+(\lambda-\varpi\deriv\vx)\sin(2z)\Big)
e^{2i\vy}&
-\varpi\deriv\vy-(\lambda-\varpi\deriv\vx)\cos(2z)\cr}
\nonumber
\eea
where $\deriv{f}$ denotes the derivative with respect to $\theta$.

We will consider two different models with the same eigenvalues but
with different eigenfunctions. We remark that since the adiabatic
theorem depends only on the properties of the eigenvalues (and
regularity properties of the projectors), it gives the same upper
bound for the correction for all the models (\ref{Kdef}) with equal
spectrum. But the actual deviation can be very different from this
estimate, depending on the efficiency of the couplings.

We can choose for example the following eigenvalues:
\be\label{model}
  \lambda_{\pm,k}(\varpi)=k\,\varpi
  \pm(\eta(\varpi)+\varpi)/2\,,\qquad\quad\mbox{where }\eta(\varpi)=
  \sqrt{(\varpi-\omega_0)^2+\Omega^2}
\ee
and $\omega_0$, $\Omega$ are constants. 
The first model is defined by choosing $x(\theta)=0$,
$y(\theta)=\theta$, i.e. $2\vy(\theta)=2\vx(\theta)=\theta$ and
$\cos(2\,z)=-(\varpi-\omega_0)/\eta(\varpi)$,
$\sin(2\,z)=\Omega/\eta(\varpi)$, hence $z$ is independent of
$\theta$.
The corresponding Floquet Hamiltonian is given by
$$
K_{\mbox{\tiny RWA}}(\theta)=-i\,\varpi\,{\partial\over\partial\theta}
+{1\over2}\pmatrix{
\omega_0&\Omega\,e^{-i\,\theta}\cr
\Omega\,e^{i\,\theta}&-\omega_0\cr
}
$$
which is the usual RWA model widely used in quantum optics.\\
The second model is defined by the choice 
$x(\theta)=-\varrho(\theta)/2$, $y(\theta)=\theta-\varrho(\theta)/2$,
i.e. $2\vy(\theta)=\theta$, $2\vx(\theta)=\theta-\varrho(\theta)$ and
the same $z$ as for the RWA case. This leads to
$$
K_{\mbox{\tiny M}}(\theta)=K_{\mbox{\tiny RWA}}(\theta)+
{\varpi\over2\,\eta}\,\deriv\varrho\,
\pmatrix{
\omega_0-\varpi&\Omega\,e^{-i\,\theta}\cr
\Omega\,e^{i\,\theta}&\varpi-\omega_0\cr
}
$$
This Hamiltonian can thus be understood as a perturbation of the
RWA model.

We consider now an extra smooth slow time dependence in the parameter
$\varpi=\varpi(s)$ and in $\varrho=\varrho(\theta,s)$. This implies
that the eigenvalues, the eigenvectors and the corresponding
eigenprojectors are smooth functions of $s$, so that the
regularity Hypothesis H0 i) is satisfied. We show in appendix that
H0 ii) is satisfied as well for any choice of smooth functions
$x, y, z $ and $\lambda$. 

We assume, for simplicity,
that $\varpi(s)=s$ (but any other smooth monotonic function of
$s$ would equally do). We select the eigenvalue
$\lambda(s)=\lambda_{+,0}(s)=(\eta(s)+s)/2$ and denote by
$\psi$ the associated eigenvector.
The only crossings that 
$\lambda$ experiences are with the $\lambda_{-,k+1}$'s and they take
place at times $s$ such that 
\be\label{crois}
  \eta(s)=k\,s, \qquad\quad k\in\Z^*.
\ee
\begin{figure}[h]
\hfil\scalebox{1.2}{\includegraphics{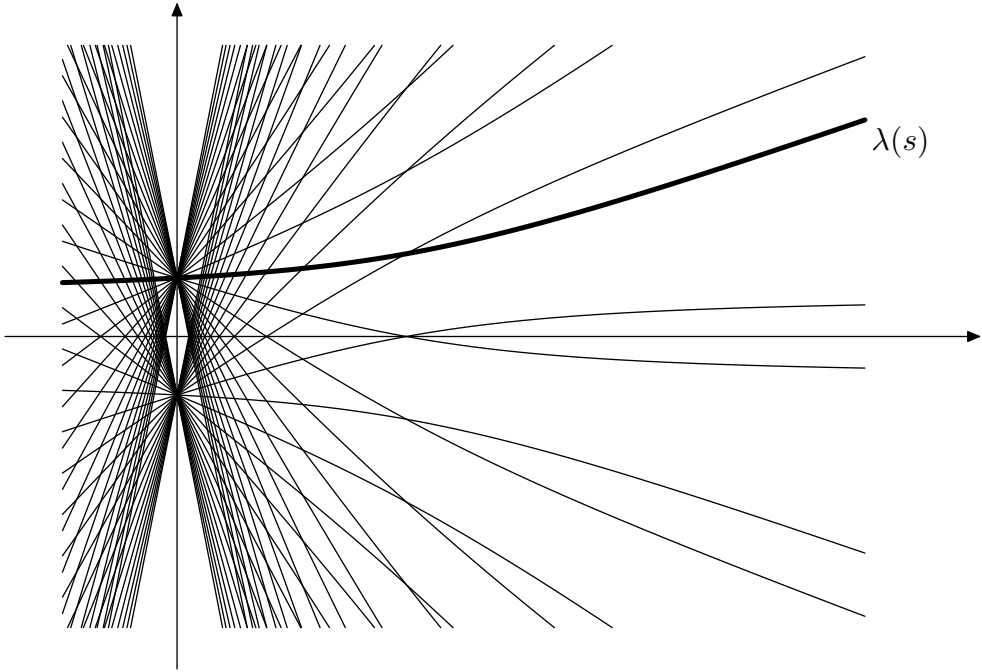}}\hfil
\caption{The first eigenvalues of the RWA and modified RWA models.}
\end{figure}
We remark however that these crossings can lead to corrections to
adiabaticity, or not, depending on whether the corresponding
eigenvectors are coupled. The non adiabatic coupling among the
branches is measured by the following scalar product:
\bea
& &\!\!\!\!\langle\,\psi(s)\mid\partial_s\psi_{-,k+1}(s)\,\rangle=
-{1\over2\pi}\int_0^{2\pi}e^{i\,(k+1)\,\theta-2\,i\,\vx(\theta,s)}
\Big(z'(s)-i\,\sin(2\,z(s))\,\vy'(\theta,s)\Big)\>d\theta\nonumber\cr
& &\phantom{\!\!\!\!\langle\psi(s)\mid\partial_s\psi_{-,k+1}(s)\rangle}=
-{z'(s)\over2\pi}\int_0^{2\pi}
e^{i\,k\,\theta+i\,\varrho(\theta,s)}\>d\theta\cr
\eea
where the $'$ denotes the derivative with respect to $s$.

Recall that the couplings between the eigenstate $\psi(s)$ associated
with the level $\lambda(s)$ and its orthogonal complement in the
Hilbert space is given by the operator $L(s)=i\,[P'(s),P(s)]$,
see (\ref{Wdiffeq}), since the adiabatic evolution $A(s)$ follows the
instantaneous eigenspaces. A direct computation of the matrix elements
$\langle\,\psi_{-,k+1}(s)\mid L(s)\,\psi(s)\,\rangle$ with
$P'(s)=\mid\psi'(s)\,\rangle\langle\psi(s)\mid
+\mid\psi(s)\,\rangle\langle\,\psi'(s)\mid$ shows that the above scalar
product is proportionnal to the couplings responsible for the non
adiabatic transitions.
\\
For the RWA model, as $\varrho=0$ the non adiabatic couplings are 
given by:
$$
\langle\,\psi(s)\mid\partial_s\psi_{-,k+1}(s)\,\rangle
=-z'(s)\,\delta_{k,0}
$$
Thus, the level $\lambda(s)$ is {\it not}  coupled to the
infinitely many other levels it crosses.  Hence we are lead in this
case to an effective  problem displaying no crossing, so that 
the error is of order $\eps$ in this case.

For the other model, we will obtain non zero couplings at all the
crossings, if we choose $\varrho(\theta,s)$ such that
$\exp(i\,\varrho(\theta,s))$ has infinitely many non zero
Fourier components. For example, one can take
$\varrho(\theta,s)=\rho(s)\,\sin(\theta)$ (in particular $\rho$ can
be chosen constant). This coupling is then given by
$$
\langle\,\psi(s)\mid\partial_s\psi_{-,k+1}(s)\,\rangle=
(-1)^{k+1}\,z'(s)\,J_k(\rho(s))
$$
where $J_k$ is a Bessel function.

We will now verify that the assumptions of 
Proposition~\ref{behaveProp} are satisfied. Let us focus on the
interval $(0,S]$, for $S$ small enough. The interval $[-S,0)$
can be treated similarly. Again to simplify the notations we will
not explicit the $+$ sub/super-scripts. 

In fact the following analysis is valid for all the models 
(\ref{Kdef}) under the assumption that the eigenvalues can be written
as $\lambda_{\pm,m}(s)=m\,s\pm\aleph(s)/2$, where $\aleph$ is a
 $C^2$ function with bounded derivatives such that
$\aleph(0)>0$. In particular they are satisfied for the
eigenvalues given in (\ref{model}).
The Hypotheses imply that the function $f_\zeta(s)=\aleph(s)-\zeta\,s$
is strictly decreasing for any $\zeta$ greater than, say, some
$\zeta_0$. Under these conditions the following assertion shows that
the crossings that $\lambda(s)=\aleph(s)/2$ experiences with the rest
of the spectrum take place at times such that $\aleph(s)=k\,s$, $k\in\N$
large enough.

\begin{asser}\label{zobehave}
For $\zeta\ge\zeta_0$, the function $f_\zeta(s)=\aleph(s)-\zeta\,s$ has a
unique positive zero $\zo_\zeta$ and if $\zeta<\xi$ we have 
$\zo_\zeta>\zo_\xi$.
\end{asser}

From the expansion
$$
f_\zeta(s)=\aleph(0)+(\aleph'(0)-\zeta)\,s+O(s^2)\,,
$$
we obtain the behaviour of $\zo_\zeta$:
\be\label{zopos}
\zo_\zeta={\aleph(0)\over\zeta-\aleph'(0)}+O(1/\zeta^3).
\ee
We define the sequence $\bd_k>0$ by the equation:
\be
\aleph(\bd_k)-k\,\bd_k=(k+1)\,\bd_k-\aleph(\bd_k)\qquad\quad\mbox{i.e. }
\aleph(\bd_k)=(k+1/2)\,\bd_k\,.
\ee
Assertion \ref{zobehave} implies that $\bd_k<\zo_k<\bd_{k-1}$ and, from
Equation (ref{zopos}) and the fact that $u_k=o_{k+1/2}$, we obtain
\be\label{compuk}
 \bd_k={\aleph(0)\over{k+1/2}-\aleph'(0)}+O(1/k^3).
\ee 
Next, we
have
\begin{asser}\label{gapass}
On the interval $[\bd_k,\bd_{k-1}]$, the spectral gap is given by
$$
g(s)=\dist\big(\lambda(s),\sigma(s)\backslash{\{\lambda(s)\}}\big)
=\big|\aleph(s)-k\,s\big|\le\bd_{k-1}/2\,.
$$
More precisely, for $\bd_k\le s\le\zo_k$ we have that
$g(s)=\aleph(s)-k\,s\le\bd_k/2$ and for $\zo_k\le s\le\bd_{k-1}$ we have
that $g(s)=k\,s-\aleph(s)\le\bd_{k-1}/2$.
\end{asser}
This assertion is easily proven by considering the different cases.\\
We now prove that the Spectral Hypothesis H1--H2 are verified.
Assertion \ref{zobehave} and Equation (\ref{compuk}) show that the
sequence $\{\bd_k\}$ is (for $k$ large enough) monotonically 
decreasing to $a=0$. To define
the intervals $V_k$, we choose any point $\ei_k$ in
$(\zo_k,\bd_{k-1})$ such that 
$g(\ei_k)=k\,\ei_k-\aleph(\ei_k)\le\bd_k/2$ and set
$V_k=(\bd_k,\ei_k)$. The $V_k$'s are disjoint and
$I_k=\{\bd_k\}\cup[\ei_k,\bd_{k-1}]$. By definition of $V_k$, we have
that $g(s)\le\bd_k/2=g(\bd_k)$ and for 
$\ei_k\le s\le\bd_{k-1}$ the gap is given by 
$g(s)=k\,s-\aleph(s)\ge\bd_k/2$. Whence, Hypothesis H1 is satisfied. 
Finally to prove that H2 holds, we need to estimate the behaviour of $g(s)$
on $V_k$: 
the Mean Value Theorem 
implies that for each $s\in V_k\backslash{\{\zo_k\}}$, there is an
$q_s$, in the interval 
joining $s$ and $\zo_k$, such that
$$
g(s)=\big|\aleph(s)-k\,s\big|=\big|k-\aleph'(q_s)\big|\>|s-\zo_k|\behave
k\,|s-\zo_k|\,,
$$
which shows that H2 is satisfied with $\alpha=1$ and $G(k)\behave k$.

It remains to check the conditions given in the statement of 
Proposition \ref{behaveProp}. We have
\be
\begin{array}{ll}
 |\bd_k-0|=\bd_k=\aleph(0)/k+\aleph(0)(\aleph'(0)-1/2)/k^2+O(1/k^3)&\qquad\mbox{i.e. }\beta=1\,\\
 G(k)\behave k &\qquad\mbox{i.e. }\gamma=1\,,\phantom{\bigg)}\\
 |V_k|\behave1/k^2&\qquad\mbox{i.e. }\delta=2\,.
\end{array}
\ee
To get the estimate for $|V_k|$, we have used that
$(\bd_k,\zo_k]\subset V_k\subset(\bd_k,\bd_{k-1}]$ and the
expressions for
$\zo_k$, and $\bd_k$ in Equations (\ref{zopos}) and (\ref{compuk}).
This implies that, $\mu=\alpha\,(\beta+1)+\gamma=1+2\,\alpha$ and
$\delta=\beta+1$. So, we can use the second
case of Proposition \ref{behaveProp} to prove that the
adiabatic approximation holds for the models:
\be
  \|U(1)-A(1)\|\le c\,\eps^p,\qquad\mbox{for any }p<\frac{1}{3}\,.
\ee
In keeping with the last remark of Section 2, we recall that
a more careful analysis yields $p=1/3$.

\appendix
\section{Appendix}

In this appendix, we show that an operator on $L^2(S^1,{\cal H})$ of the form
\be\label{Ks}
K(s,\theta)=-i\,\varpi(s)\,{\partial\over\partial\theta} 
             +H(s,\theta)\,,
\ee
where $H(s,\theta)$ is a bounded operator in ${\cal H}$ such that 
$s\mapsto H(s,\theta)$ and 
$s\mapsto\partial/\partial\theta H(s,\theta)$ are norm continuous
and $s\mapsto\varpi(s)$ is continuous, admits a strongly continuous
unitary propagator $U(s)=U(s,0)$ with all expected regularity
properties, even if there is a value $a$ for 
which $\varpi(a)=0$. Notice that the assumptions on $H$ will be
satisfied if, for example, $(s,\theta)\mapsto H(s,\theta)$ is strongly
$C^1$. 

The proof relies on a theorem of T. Kato \cite{k2}, which we will
restate in a more suitable form for our purpose.

\begin{thm}[Kato]\label{Kato} Let $\cK$ and $\cL$ be Hilbert spaces such
that $\cL$ is densely and continuously embedded in $\cK$ and
let $K(t)$, $0\le t\le T$, be a family of self-adjoint operators in
$\cK$. Suppose that
\begin{itemize}
\item[\it 1)] $\cL\subset\dom K(t)$ for all $0\le t\le T$,
whence the $K(t)$ are bounded operators from $\cL$ to $\cK$,
and the application $t\mapsto K(t)$ is norm continuous from
$\cL$ to $\cK$;
\item[\it 2)]there exists a family of isomorphisms $S(t)$ from 
${\cal D}$ to $\cK$ which is strongly continuously differentiable and
such that
$$
S(t)\,K(t)\,S(t)^{-1}=K(t)+B(t)
$$
where $B(t)$ is a strongly continuous bounded operator on $\cK$.
\end{itemize}
Under those conditions, there exists a unique family of unitary
operators $U(t,s)$ on $\cK$ defined for $0\le s,t\le T$ with the
following properties:
\begin{itemize}
\item[{\it i)}]$U(t,s)$ is strongly continuous on $\cK$ in $s,t$ with
$U(s,s)=\un$;
\item[{\it ii)}]$U(t,r)=U(t,s)\,U(s,r)$;
\item[{\it iii)}]$U(t,s)\,\cL\subset\cL$,
$\|U(t,s)\|_\cL\le N\,{\rm e}^{\,c\,|t-s|}$ and is strongly continuous 
on $\cL$ in $s,t$ simultaneously;
\item[{\it iv)}]$\displaystyle{d\over ds}U(t,s)\,\psi= 
i\,U(t,s)\,K(s)\,\psi$ for any $\psi\in\cL$, for $0\le s,t\le T$;
\item[{\it v)}]for each $\psi\in\cL$ and fixed $s$, 
$\displaystyle{d\over dt}U(t,s)\,\psi$ exists and is equal to 
$-i\,K(t)\,U(t,s)\,\psi$ and strongly continuous in $\cK$ in $t$.
\end{itemize}
\end{thm}

To prove this theorem, we apply Theorem 6.1 in \cite{k2} to the 
operator $A(t)=i\,K(t)$, which is stable with constants of stability 
$c=0$ and $N=1$ (see Definition 3.1 and Theorem 4.1 therein). The 
fact that $U(t,s)$ is unitary follows from the self-adjointness of 
$K(t)$, the construction of $U(t,s)$ by unitary approximants given in 
the proofs of Theorem 4.1 and 6.1 in \cite{k2} and the invertibility 
of $U(t,s)$, which is a consequence of the fact that 
$A^{^{_{\scriptstyle\circ}}}\!(t)=-i\,K(T-t)$ satifies also the 
hypothesis of Theorem 6.1 in \cite{k2}. See also Remark 5.3 therein.

We now prove that the family of self-adjoint operators defined by 
Equation (\ref{Ks}) satisfies the hypothesis of Theorem \ref{Kato}. To
simplify the notation, we will not explicit the $\theta$-dependence 
and write $\partial$  for $\partial/\partial\theta$.
\newline{\bf Proof:}\\
For $\cL$, we choose $\dom(-i\,\gmm\partial)$ for some $\gmm>0$, 
and we notice that for any $t$ such that $\varpi(t)\not=0$, we have 
that $\dom K(t)=\cL$ and if $\varpi(t)=0$, then $\dom K(t)=\cK$. For 
the norm on $\cL$, we choose the graph norm associated to 
$-i\,\gmm\partial$:
$$
\|\psi\|_\cL{}^2=\|\psi\|^2+\|-i\,\gmm\,\partial\,\psi\|^2
\ge\|\psi\|^2\,.
$$
Whence, $\cL$ is a dense continuously embedded subspace of $\cK$.
\newline
For any $s,t$ and any $\psi\in\cL$, we have
\bea
\big\|\big(K(t)-K(s)\big)\,\psi\big\|^2&\le&
2\,{|\varpi(t)-\varpi(s)|^2\over\gmm^2}\>
\|-i\,\gmm\partial\,\psi\|^2+2\,
\big\|\big(H(t)-H(s)\big)\,\psi\big\|^2\nonumber\phantom{\Bigg)}\\
&\le&2\,\max\left\{{|\varpi(t)-\varpi(s)|^2\over\gmm^2};
\big\|H(t)-H(s)\big\|^2\right\}\>\|\psi\|_\cL{}^2\,.\nonumber
\eea
which shows the norm continuity of $K(t)$.

We set $S(t)=S=-i\,\gmm\partial+i$. $S$ is an isomorphism between
$\cL$ and $\cK$ which is strongly differentiable (by $t$
independence). It remains to show that $S$ satisfies Hypothesis 2) of
Theorem \ref{Kato}.
For this, we first notice that for any $\psi\in\dom K(t)$, we have 
that $S^{-1}\psi\in\cL\subset\dom K(t)$ and
\bea\label{KS-1}
K(t)\,S^{-1}\psi&=&S^{-1}K(t)\,\psi+H(t)\,S^{-1}\psi-S^{-1}H(t)\,\psi
\nonumber\\
&=&S^{-1}K(t)\,\psi+S^{-1}S\,H(t)\,S^{-1}\psi-S^{-1}H(t)\,\psi
\nonumber\\
&=&S^{-1}\,\Big(K(t)-i\,\gmm\partial H(t)\,S^{-1}\Big)\,\psi\,.
\eea
Whence, for any $\psi\in\dom K(t)$, we have that the left-hand side
of Equation (\ref{KS-1}) belongs to $\cL$. So we can write,
$$
S\,K(t)\,S^{-1}\psi=K(t)\,\psi-i\,\gmm\partial H(t)\,S^{-1}\psi\,,
\qquad\mbox{for all }\psi\in\dom K(t)\,.
$$
Setting $B(t)=-i\,\gmm\partial H(t)\,S^{-1}$, we have a strongly 
continuous bounded operator (by the assumptions on $H$)
which satisfies $S\,K(t)\,S^{-1}\supset K(t)+B(t)$. To show the
reverse inclusion, we can consider any $b\ge2\,\sup_t\|B(t)\|$ which
implies that $i\,b$ belongs to the resolvent set of both $K(t)+B(t)$ 
and $S\,K(t)\,S^{-1}$. It follows that 
$\big(K(t)+B(t)+i\,b\big)^{-1}\subset
S\,\big(K(t)+i\,b\big)^{-1}\,S^{-1}$. But since the left hand side has
domain $\cK$, we must have equality between $K(t)+B(t)$ and 
$S\,K(t)\,S^{-1}$ instead of inclusion.
\ep

In the examples of Section 4, both $H(s,\theta)$
defined through (\ref{Kdef}) by means of smooth fonctions 
$x, y, z, \lambda$ of $(s,\theta)$, and $H(s,\theta)+\eps i 
[P'(s,\theta),P(s,\theta)]$ where $P(s,\theta)=|\psi(s,\theta)\ket
\bra \psi(s,\theta)|$ with $\psi(s,\theta)$ given by one of the vectors
(\ref{defpsi}) satisfy the hypotheses of the Theorem. Hence 
assumption H0 ii) is satisfied for these models.

\section*{Acknowledgements}
Supports from the EC contract {\sc erbchrxct}94-0460 for the project
``Stability and universality in classical mechanics'', from the
{\sc cnrs} and from grants of the Fonds National Suisse de la
Recherche Scientifique are acknowledged.


\begin{thebibliography}{XXXX}

\bibitem[AE]{ae} J.E.~Avron, A.~Elgart, {\sl The Adiabatic Theorem
of Quantum Mechanics}, {\em preprint} (june 1998).
\bibitem[AE2]{ae2} J.E.~Avron, A.~Elgart, {\sl An Adiabatic Theorem
without a Gap Condition}, to appear in {\em Proceedings of Qmath7}, 
Prague, (june 1998).
\bibitem[AHS]{ahs} J.~Avron , J.~Howland, B.~Simon, Adiabatic Theorems
for Dense Point Spectra, {\em Commun.Math.Phys.} {\bf 128} (1990) 
497-507.
\bibitem[ASY]{asy} J.E.~Avron, R.~Seiler, L.G.~Yaffe, {\sl Adiabatic
Theorems and Applications to the Quantum Hall Effect}, 
{\em Commun.\ Math.\ Phys.} {\bf 110} (1987) 33--49.
\bibitem[BF]{bf}M.~Born, V.~Fock, {\sl Beweis des Adiabatensatzes},
{\em Zeit.\ f.\ Physik} {\bf 51} (1928) 165--180.
\bibitem[GJ]{gj} S.~Gu\'erin, H.R.~Jauslin, {\sl Laser-Enhanced
Tunnelling Through Resonant Intermediate Levels},
{\em Phys.\ Rev.}
{\bf A 55} (1997) 1262--1275.
\bibitem[GMDJ]{gmdj} S.~Gu\'erin, F.~Monti, J.-M.~Dupont,
H.R.~Jauslin, {\sl On the Relation Between Cavity-Dressed States,
Floquet States, RWA and Semiclassical Models},
{\em J.~Phys.\ A: Math.\ Gen.}
{\bf 30} (1997) 7193--7215.
\bibitem[Hag]{ha0}G.~Hagedorn, {\sl Adiabatic Expansions near
Eigenvalue Crossings}, {\em Ann.Phys.} {\bf 196} (1989) 278-295.
\bibitem[Hol]{hol} M.~Holthaus, {\sl Pulse-Shape-controlled Tunnelling
in a Laser Field}, {\em Phys.\ Rev.\ Lett.}, 
{\bf 69} (1992) 1596.
\bibitem[Joy]{j2} A.~Joye, {\sl An Adiabatic Theorem for Singularly
Perturbed Hamiltonian}, {\em Ann.\ Inst.\ H.\ Poincar\'e, Sect.\ A}
{\bf 63} (1995) 231--250.
\bibitem[JP]{jp} A.~Joye, Ch.-Ed.Pfister : 
        {\sl Exponential Estimates in Adiabatic Quantum 
Evolution}, to appear in: {\em Proceedings of the XIIth International 
Congress of Mathematical Physics, Brisbane, 13-19 July 1997.}
\bibitem[Ka1]{ka1} T.~Kato, On the Adiabatic Theorem of Quantum Mechanics,
{\em J.Phys.Soc.Japan} {\bf 5} (1950) 435-439.
\bibitem[Ka2]{k2} T.~Kato, {\sl Linear Evolution Equation of 
``Hyperbolic'' Type}, {\em J.\ Fac.\ Sci.\ Univ.\ T\=oky\=o, 
Sect.\ I,A Math.} {\bf 17} (1970) 241--258.
\bibitem[Kr]{kr} S.G.~Krein, {\em Linear Differential Equations in 
Banach Space} (American Mathematical Society Providence 1971).
\bibitem[Nen]{ne} G.~Nenciu, On the Adiabatic Theorem of Quantum 
Mechanics, {\em J.Phys.A} {\bf 13} (1980) L15-L18.
\end{thebibliography}
\end{document}